\documentclass{iau} 
\usepackage{graphicx}

\title[PNLF]
{The Planetary Nebulae Luminosity Function (PNLF): current perspectives}

\author[Roberto H. M\'endez] {Roberto H. M\'endez}

\affiliation{Institute for Astronomy, University of Hawaii, \\ 2680 Woodlawn Drive, Honolulu, HI 96822, USA \\ email: {mendez@ifa.hawaii.edu}}

\pubyear{2017}
\volume{323}
\jname{Planetary Nebulae: Multi-wavelength Probes of Stellar and Galactic Evolution}
\editors{Xiaowei Liu, Letizia Stanghellini, \& Amanda Karakas, eds.}
\begin{document}

\maketitle

\begin{abstract}
This paper starts with a brief historical review about the PNLF and its use as 
a distance indicator. Then the PNLF distances are compared with Surface Brightness Fluctuations (SBF) distances and Tip of the Red Giant Branch (TRGB) distances. A Monte Carlo method to generate simulated PNLFs is described, leading to the last subject: recent progress in reproducing the expected maximum final mass in old stellar populations, a stellar astrophysics enigma that has been challenging us for quite some time.
\end{abstract}

\firstsection 

\section{Historical introduction}

The field of extragalactic PNs as tools for distance determination was pioneered by Holland Ford and George Jacoby, later joined by Robin Ciardullo. They started detecting and measuring PNs in nearby galaxies in the late 70's and early 80's. The measured fluxes in [O III] 5007 can be transformed into ``Jacoby magnitudes'' $m$(5007) (Jacoby 1989). The PNLF then indicates how many PNs are found at each apparent magnitude $m$(5007). It was noticed that the absolute magnitudes $M$(5007) of the brightest PNs in galaxies of known distance always had a similar value. On this mostly empirical basis, they proposed to assume that the bright end of the PNLF is universal, and use it as a standard candle. An analytical representation of the PNLF was introduced by Ciardullo et al. (1989). The absolute magnitude $M^*$ of the bright end needs to be calibrated using a galaxy at a known distance, for example the bulge of M 31.
After testing the method on other nearby galaxies, Jacoby et al. (1990) applied it to several galaxies in the Virgo cluster, obtaining similar distances to all those galaxies, and an average distance of 15 Mpc, implying a high value of the Hubble constant that was later essentially confirmed by the HST Key project on cepheid distances. As of today, the PNLF method has given distances to about 60 galaxies. 


\section{The discrepancy with SBF distances}

In the meantime, other methods of distance determination have become popular, among them the SBF method (Tonry and Schneider 1988, Tonry et al. 2001, Blakeslee et al. 2009). Around the year 2000, it became apparent that there is a systematic difference between PNLF and SBF distances, with PNLF distance moduli being smaller by about 0.3-0.4 mag at distances in excess of 10 Mpc (M\'endez 1999; Ciardullo et al. 2002). A more recent comparison adding a few more galaxies is shown in Fig. 16 of Teodorescu et al. (2010).

I have no room to describe ideas about the reason for the discrepancy (see e.g. Ciardullo et al. 2002, and particularly the careful review by Ciardullo 2012);
but it is important to show what happens if we assume that the SBF distances are correct.
Figure 1 shows how the increased distance forces the observed PNLF to become more luminous. The brightest PNs become too bright. Why don't we see such superbright PNs in nearby galaxies? If SBF distances are confirmed, this will require an explanation.

\begin{figure}[b]
\begin{center}
 \includegraphics[width=3.4in]{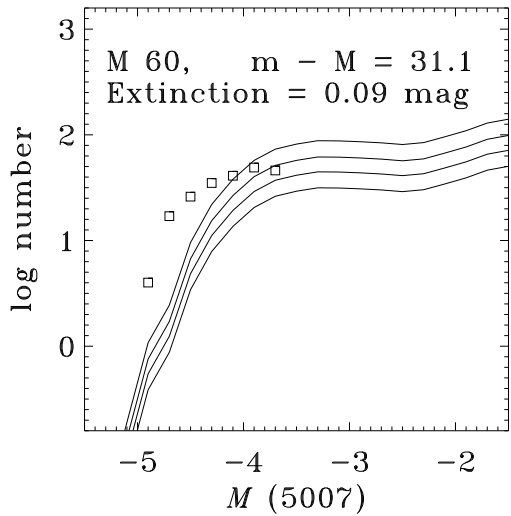} 
 \caption{Observed [O III] $\lambda$ 5007 PNLF of M 60 (squares). The PNLF distance to M 60 corresponds to a distance modulus 30.7 (Teodorescu et al. 2011). Here I have forced this observed PNLF to be at the SBF distance modulus 31.1 (Blakeslee et al. 2009). Therefore all the PNs have become 0.4 mag brighter. The simulated PNLF (solid lines) is the same one used by Teodorescu et al. (2011).}
   \label{fig1}
\end{center}
\end{figure}

\section{TRGB distances as a tie-breaker}

We need a third method to decide if it is the PNLF distances or the SBF distances, or even both, that are systematically wrong. I believe our best choice for this test is TRGB distances. The method is described by Makarov et al. (2006); it has been applied to many galaxies up to distances of about 10 Mpc. The agreement with PNLF distances is good (Ciardullo 2012). Unfortunately, right now we have only one TRGB distance to a galaxy more distant than 14 Mpc: M 87 (Bird et al. 2010). This single TRGB distance agrees with the SBF distance. Ideally, we want at least ten TRGB distances in the Virgo and Fornax clusters to yield a convincing result and decide once and for all what is the truth. This will be possible with the James Webb Space Telescope (JWST), which is expected to allow TRGB distance determinations up to at least 30 Mpc.

\section{Modeling the PNLF} 

It is not necessary to rely exclusively on the analytical PNLF representation. An alternative is to produce simulated PNLFs, using Monte Carlo techniques. This approach is described by M\'endez et al. (1993), M\'endez \& Soffner (1997), and M\'endez et al. (2008).
The idea is to generate a set of PN central stars with random post-AGB ages and masses.
Using a family of post-AGB evolutionary tracks, for every pair (age, mass) the corresponding central star luminosity and surface temperature can be derived.
Using recombination theory and empirical information about leaking of H-ionizing photons, the nebular $H\beta$ luminosities are calculated. Finally, generating suitable ratios $I(5007)/I(H\beta)$ for different regions of the HR diagram, the luminosities in 5007 are obtained, and the PNLF is built.

\section{The mass distribution and the maximum final mass enigma}

In such PNLF simulation work we have always used a mass distribution similar to that of white dwarfs. For a population without recent star formation, it is necessary to cut the mass distribution at a certain maximum final mass, because all the more massive stars have already evolved into white dwarfs. Here comes the problem: since we expect elliptical galaxies to have old populations, the initial mass is more or less 1 $M_\odot$, and therefore the maximum final mass is expected to be at most 0.55 $M_\odot$. The PNLF should become very faint. However, the observed PNLF bright end stays constant. From the observed central star luminosities, using traditional post-AGB evolutionary tracks, the maximum final mass is predicted to be ~0.63 $M_\odot$; a severe inconsistency.

\section{New evolutionary tracks and their consequences}

The old tracks required a central star mass of 0.63 $M_\odot$ to produce 
log $L/L_{\odot}$ = 3.9
as the star abandons the AGB. New evolutionary tracks by Miller Bertolami (2016) have considerably reduced the required central star mass. Figure 2 shows a fit to the PNLF of the elliptical galaxy NGC 4697 (M\'endez et al. 2001), produced with a simulation based on an analytical approximation to Miller Bertolami's evolutionary tracks. Now the mass distribution can be truncated at 0.58 $M_\odot$. This is perhaps not a full solution to the maximum final mass enigma, but it is certainly a substantial step in the right direction. With a little help from the initial-final mass relation as described by Cummings in these Proceedings, we may be close to solving this persistent problem.

\begin{figure}[b]
\begin{center}
 \includegraphics[width=3.4in]{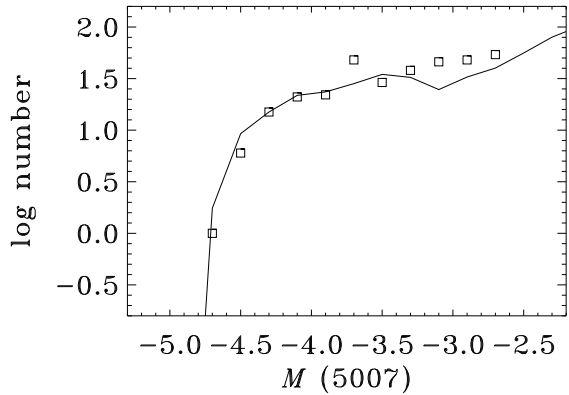} 
 \caption{Observed PNLF of NGC 4697 (squares; distance modulus 30.1; from Fig. 14 of M\'endez et al. 2001), compared with a PNLF simulation produced as described in Section 6. Now the maximum final mass can be reduced to 0.58 $M_\odot$, in much better agreement with the initial-final mass relation (see Cummings, these Proceedings).}
\label{fig1}
\end{center}
\end{figure}

\section{PN formation rates}

There is another important consequence of the new Miller Bertolami tracks. In addition to being more luminous at any given mass, these more luminous central stars move faster along their tracks; there is less available fuel, and it is consumed more quickly. The times required to cross the HR diagram at high luminosity are shorter.

It is customary to define a formation rate $B_{\rm PN}$, in PNs per year per $L_\odot$; 
a PN visibility duration $t_{\rm PN}$ in years (e.g. 30,000); the total luminosity
$L_{\rm T}$ of the sampled population in $L_\odot$; and the total PN population 
$n_{\rm PN}$ (only a fraction of these -the brightest- are actually observed). Then we can write $ n_{\rm PN} = B_{\rm PN} L_{\rm T} t_{\rm PN} $. The luminosity-specific PN formation rate $\alpha$ in PNs per $L_{\odot}$ is given 
by $ \alpha = B_{\rm PN} t_{\rm PN} = n_{\rm PN} / L_{\rm T} $.

Now introduce the new evolutionary tracks, with their shorter crossing times. The percentage of visible PNs goes down, because we have not modified $t_{\rm PN}$. So
from a given number of observed PNs we derive a larger $n_{\rm PN}$. Consequently
$B_{\rm PN}$ and $\alpha$ increase. The new values of $B_{\rm PN}$ and $\alpha$ may
turn out to be in better agreement with the predictions of simple stellar population (SSP) theory (Renzini \& Buzzoni 1986; Buzzoni et al. 2006).

\section{A view to the future}

\noindent 1. We may be on our way to finally solving the maximum final mass enigma (Miller Bertolami's new evolutionary tracks, plus a little help from the initial-final mass relation).

\noindent 2. The PNLF cannot be expected to play a dominant role in future extragalactic distance determination; other methods have advantages. For example, the PNLF method cannot migrate toward the infrared, like the SBF and TRGB methods. 

\noindent 3. On the other hand, once the PNLF-SBF distance problem is fixed, we will gain confidence to use the PNLF and consequent limits on PN formation rates to learn about post-AGB evolution in a variety of stellar populations.

\end{document}